# Stokes' Cradle: Normal Three-Body Collisions between Wetted Particles

By C. M. Donahue, C. M. Hrenya[†], R. H. Davis, K. J. Nakagawa, A. P. Zelinskaya, and G. G. Joseph

Department of Chemical and Biological Engineering, University of Colorado, Boulder, CO 80309–0424, USA

In this work, a combination of experiments and theory is used to investigate three-body, normal collisions between solid particles with a liquid coating (i.e., "wetted" particles). Experiments are carried out using a Stokes' cradle, an apparatus inspired by the Newton's cradle desktop toy except with wetted particles. Unlike previous work on two-body systems, which may either agglomerate or rebound upon collision, four outcomes are possible in three-body systems: fully agglomerated, Newton's cradle (striker and target particle it strikes agglomerate), reverse Newton's cradle (targets agglomerate while striker separates), and fully separated. Post-collisional velocities are measured over a range of parameters. For all experiments, as the impact velocity increases, the progression of outcomes observed is fully agglomerated, reverse Newton's cradle, and fully separated. Notably, as the viscosity of the oil increases, experiments reveal a decrease in the critical Stokes number (the Stokes number that demarcates a transition from agglomeration to separation) for both sets of adjacent particles. A scaling theory is developed based on lubrication forces and particle deformation and elasticity. Unlike previous work for two-particle systems, two pieces of physics are found to be critical in the prediction of a regime map that is consistent with experiments: (i) an additional resistance upon rebound of the target particles due to the pre-existing liquid bridge between them (which has no counterpart in two-particle collisions), and (ii) the addition of a rebound criterion due to glass transition of the liquid layer at high pressure between colliding particles.

## 1. Introduction

Granular flows occur in diverse physical systems—from corn in a hopper to the spray of soil from lunar landings. A wide amount of interest exists in modeling collisions between solid particles as a path to unlock the overall flow dynamics. Among the earliest research efforts into contact mechanics was by Hertz (1882) over a century ago as he strove to understand the optical properties of stacked lenses with the assumptions of frictionless materials and perfect elasticity. Since then, advances have been made to describe additional complexities of such collisions, for example, their surface roughness and dissipative nature. In this work, the added complexity of liquid-coated solid particles (i.e., "wetted" particles) undergoing collisions is studied. Such wetted flows are found in numerous settings in both industry and nature. Chemical and pharmaceutical industries incorporate wetted particles in processes such as filtration, granulation, spray coating, pneumatic transport and coagulation. Natural processes involving wetted particles include pollen capture, avalanches, and sedimentation.

---

[†] Author to whom correspondence should be addressed: hrenya@colorado.edu

To date, all studies of wetted collisions have focused on two-body systems, in which the only two possible outcomes are agglomeration or separation. In this effort, the focus is on three-body collisions between an incoming striker particle and two initially touching, motionless, target particles (i.e., these particles are initially agglomerated), with all particles arranged in a line to ensure normal collisions. With the addition of this third particle, now four outcomes are possible for wetted systems: fully agglomerated (FA); "Newton's cradle" (NC), in which the striker and the target particle it strikes agglomerate while the last target particle is separated, named after the outcome commonly associated with the (dry) desktop toy; "reverse Newton's cradle" (RNC), in which the striker is separated and the two targets are agglomerated; and fully separated (FS). In this work, the focus is restricted to wetted collisions characterized by low Reynolds number (ratio of fluid inertia to fluid viscous forces in the liquid gap between colliding particles), $Re$, and high capillary number (ratio of fluid viscous forces to capillary forces in the gap), $Ca$. The foundation of the description for such wetted collisions traces to earlier work on immersed collisions between two particles. The Stokes number,

$$St = \frac{\tilde{m} v_0}{6\pi \mu \tilde{a}^2},$$ (1)

which is a measure of the inertia of colliding particles relative to the viscous force of the surrounding liquid, is the relevant dimensionless number. Here, $\tilde{m}$ is the reduced mass of the particles ($\tilde{m} = m_1 m_2/(m_1+m_2)$, where subscripts indicate different particles), $v_0$ is the initial relative velocity between the two particles, $\mu$ is the viscosity of the liquid, and $\tilde{a}$ is the reduced radius of the particles ($\tilde{a} = a_1 a_2/(a_1+a_2)$). Low-Reynolds-number (lubrication) theory has established that two smooth, rigid particles approaching one another will never touch or rebound, but instead stop at a finite distance as they approach. The deformation of immersed (non-rigid) particles was first considered by Davis, Serayssol & Hinch (1986). In their work, a model was developed which couples the fluid hydrodynamics and the particle (elastic) deformation during the collision, known as elastohydrodynamics. In this manner, kinetic energy is stored in the deformation and, when it is released, rebound of the particle may be achieved depending on the $St$. Additionally, their predictions indicate that, as the viscosity of the fluid increases, the critical Stokes number, $St_c$, decreases, where $St_c$ is the Stokes number at which there is a transition from agglomeration to rebound. Later work by Barnocky & Davis (1989) includes a pressure-dependent viscosity proposed by Chu & Cameron (1962) in their model of a two-body immersed collision. Barnocky & Davis (1989) concluded that, while the inclusion of pressure-dependent viscosity lowers the $St_c$, it plays a weak role in the outcomes of the collision in their parameter space. Experimental collisions performed by measuring the velocity of a particle as it bounced off a wall immersed in liquid confirm the described theoretical trends (Joseph *et al.*, 2001).

Numerous investigations have also been performed for wetted two-body collisions, expanding on the aforementioned works on immersed collisions. Some of the earlier experimental works that consider wetted-particle collisions include Barnocky & Davis (1988) and Lundberg & Shen (1992), who performed two-body collisions by dropping dry particles onto a liquid-coated surface. Both works confirm the trend proposed by Davis *et al.* (1986)(1986) of decreasing $St_c$ with increasing viscosity. Ennis, Tardos & Pfeffer (1991) modeled a two-body wetted collision without employing elastohydrodynamics, in order to make relative conclusions about the granulation process.  Their model is unable to predict the $St_c$ trends with a change in viscosity, which will be explained below.  Lian, Adams & Thornton (1993) presented a slightly simplified model of wetted collisions based on elastohydrodrynamics that agrees well with Davis *et al.* (1986).  In the effort by Davis, Rager & Good (2002), a scaling argument was used to apply the elastohydrodynamic theory developed by Davis *et al.* (1986) to wetted particles.  Further work by Kantak & Davis (2006) presented a complete elastohydrodynamic coupling to describe wetted collisions.

To build on previous efforts, the focus of the current effort is on normal (head-on), three-body, wetted collisions, which are investigated using a combination of experiments and theory.  The experiments are conducted using an apparatus inspired by the Newton's cradle desktop toy.  In this "wetted" operation, the apparatus is referred to as the "Stokes' cradle" since the fluid motion in the liquid layer between colliding particles is described by Stokes (low *Re*) flow.  A series of experiments is conducted with variations in fluid viscosity, thickness of the liquid layer, particle material, and impact velocity of the striker particle.  Comparisons of observed outcomes to predictions reveal new and interesting physical processes not present in to two-body systems.  First, for a three-particle collision, excess liquid exists in the bridge connecting the two initially agglomerated target particles (whereas two-particle collisions do not have a liquid bridge prior to contact).  Because the thickness of this bridge is orders of magnitude larger during the rebound phase compared to the initial liquid thickness between target particles, the additional resistance provided by this "excess" liquid is key to capturing the outcomes observed experimentally.  Second, the glass transition of the liquid layer between colliding particles adds more "bounce", which proves to be essential in predicting the correct outcomes.

## 2. Experimental Setup, Materials and Methods

The Stokes' cradle is created from three pairs of hanging, V-shaped pendulum arms as illustrated in figure 1.  The pivot points of each pendulum are separated by approximately 33 cm, and the length of each line from the suspension apparatus to the particles is 1 m.  The three pendulums are spaced 2.9 cm apart, which is slightly larger than one particle diameter (2.54 cm).  This extra spacing ensures that sufficient space exists for a liquid layer of non-zero thickness (i.e., liquid bridge) between the two motionless target particles at the bottom of the arc; if the pendulums were placed one diameter apart, the surfaces of the two particles would touch.

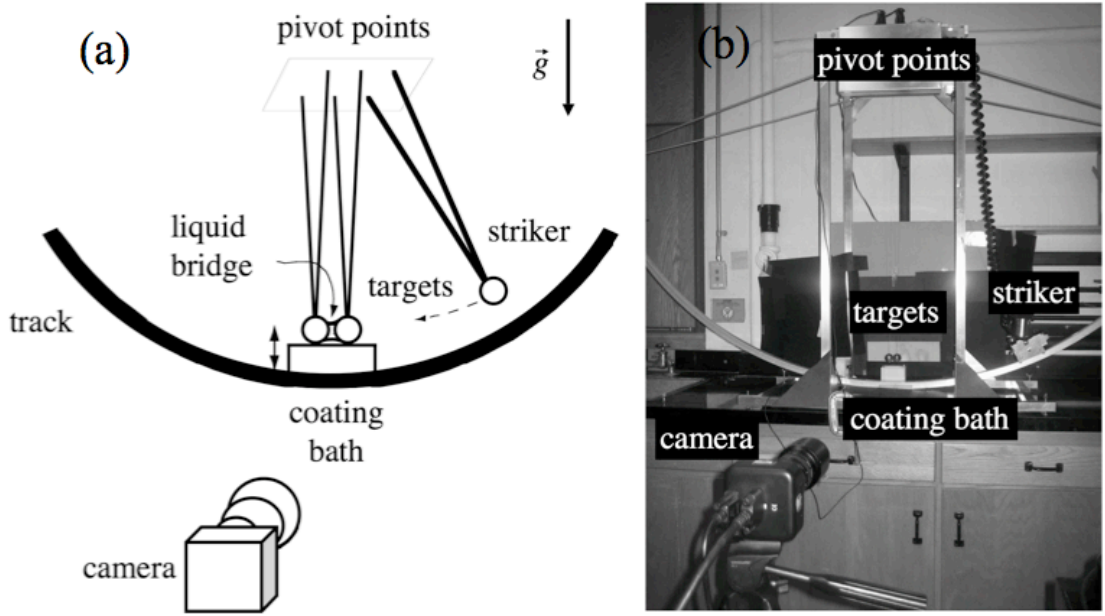

FIGURE 1. (a) Schematic and (b) photograph of Stokes' cradle experimental setup.

## 2.1. *Materials*

The pendulum lines are made of ice fishing line manufactured by Berkley, chosen for its high spring constant of 1.2 N/m. The stiff line balances the centripetal force experienced by the striker particle as it is released and travels down the arc, effectively eliminating any upward motion upon collision with the stationary particles at the bottom of the arc. The line is attached to the particles via a small, metal tube welded on the top of the particles, and all-purpose glue holds the line and tube together. For a given experiment, all three particles are fabricated from one of two types of materials, chrome steel (AISI 52100) or stainless steel (316 grade). The properties of the chrome-steel particles are: dry restitution coefficient $e_d = 0.99$; Young's modulus $E = 2.03 \times 10^{11}$ N/m$^2$; Poisson's ratio $\nu = 0.28$; density $\rho = 7830$ kg/m$^3$; radius $a = 1.27$ cm. The properties of the stainless-steel particles are: dry restitution coefficient $e_d = 0.9$; Young's modulus $E = 1.93 \times 10^{11}$ N/m$^2$; Poisson's ratio $\nu = 0.35$; density $\rho = 8030$ kg/m$^3$; radius $a = 1.27$ cm. Two silicon oils with different viscosities are used to coat the particles, namely 12000 cP and 5120 cP at 25 °C, the nominal temperature of the experiments. The oil densities are both 0.97 g/cm$^3$.

## 2.2. *Methods*

Example snapshots taken during the collision process are shown in figure 2. Particle 1 refers to the striker particle, particle 2 refers to the first target particle, and particle 3 is the end target particle opposite to the striker particle. Two types of measurements are taken to characterize each series of collisions: (i) the initial thickness of the liquid layers between the two target particles, $x_{0,2\text{-}3}$, and between the striker/target particles, $x_{0,1\text{-}2}$ (figure 3 (b)), and (ii) the pre- and post-collisional velocities of each particle after the first series (right-to-left) of collisions. As detailed below, the former is performed off-line with a high-resolution camera, while the latter is performed with a separate high-speed camera.

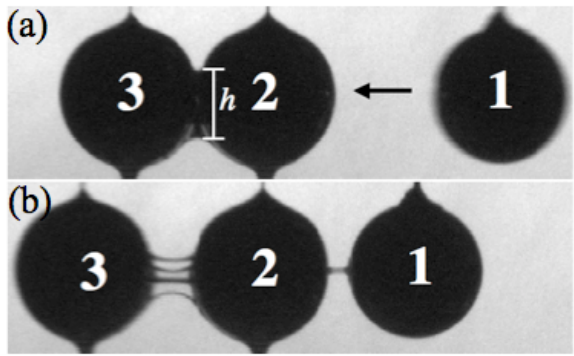

FIGURE 2. Snapshots of a three-particle wetted collision (a) just prior to collision and (b) after the collision using 12000 cP oil viscosity and stainless-steel balls (case lμ_ss_tn in table 2).

At the beginning of the liquid-layer measurements, the two target particles are wetted using a coating bath placed underneath the particles, as shown in figure 3 (a). The coating bath is raised to immerse the particles in silicon oil and is then slowly lowered. The thickness of the layer will vary with time as the silicon oil drips from the particle. Accordingly, the oil thickness is measured over a range of time. Measurements of the oil thicknesses are made via a high-resolution camera, a Pentax SLR K110D with 6.1 megapixels. To minimize the effect of wide-angle distortion, a zoom lens is used so that the camera can be placed approximately 1.5 m away from the pendulum apparatus. Photographs of the wet particles are taken every 3 seconds during the dripping process. Figure 3 (b) is a representative photograph used to calculate the liquid thickness. The lighting, aperture, and shutter speed are set at levels to make the particle, and particularly the edge of the particle, well defined and dark with respect to the background. The particles are almost entirely darker than the background (except for where the flash is reflected) and at the top of each particle the green dots contrast against the red background (though not apparent from the black and white photograph). The dots serve as a reference point for image processing using built-in Matlab functions. Matlab analysis also locates the position of the outermost edge of the particles, which is the initial point of contact during the collision. Furthermore, photographs of the dry particles are also taken prior to their wetting. From these positions in the dry and wet photographs, the geometry of the particle positions is sufficiently defined and the thickness of the outer layer, $x_{0,1-2}$, and the thickness of the inner layer between the particles, $x_{0,2-3}$, can be calculated (figure 3 (b)). An example of the dependence of the layer thicknesses with time is shown in figure 3 (c).

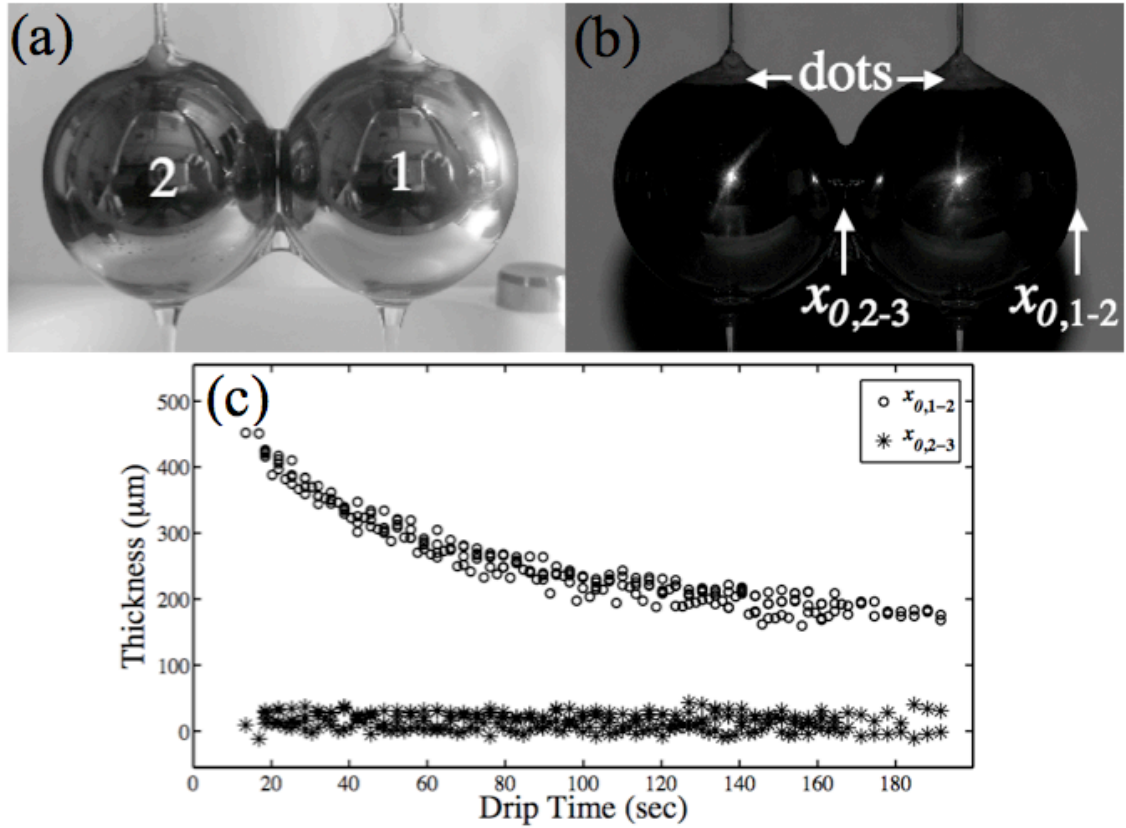

FIGURE 3. (a) Photograph of the target particles during the dripping process and (b) high-contrast snapshot taken with the Pentax high-resolution camera. (c) Plot of the thickness versus time for 5120 cP oil viscosity and stainless-steel balls.

It is important to note that, when the particles are wetted, the surface tension associated with the liquid bridge pulls the particles together. Therefore, the pendulum arms move a small angle toward one another. Even though this angle is quite small (~0.1°), its influence on the measurement of the oil layer thicknesses $x_{0,1-2}$ and $x_{0,2-3}$ is non-negligible and thus is taken into account when calculating the thicknesses. The error of $x_{0,2-3}$ measurements is relatively large, considering that a few negative thicknesses are calculated. To verify the measurements of $x_{0,2-3}$, a small spacer with a known thickness (100 – 315 μm) is placed in between the two target particles while they are dry. The thickness of the spacer is calculated using the methods described above and compared to the known thickness, resulting in an error on the order of 10 μm. Although the error is comparable to the size of $x_{0,2-3}$, predictions from the model presented later do not qualitatively change when $x_{0,2-3}$ is set equal to the size of the surface roughness (lower bound of $x_{0,2-3}$) and when the error of 10 μm has been added to the averaged $x_{0,2-3}$ (upper bound). Therefore, the error associated with the measurement of $x_{0,2-3}$ does not change the conclusions of this work.

Once the oil-layer thicknesses are established as described above, the collisional measurements are carried out. Again, the two dry target particles are dipped in the coating bath, and the time at which the collision is carried out is based on the desired oil thickness for that measurement as established previously (for example, using linear fit of the data in figure 3, five seconds before and after the collision time). The striker particle is not coated, but since it is impacting a wet target (particle 2), the collision between the two is wetted – i.e., there is a liquid layer in between the particles. The normal, three-body collision is achieved by pulling back along the arc the (dry) striker particle, which is then released and allowed to collide with the two motionless, wetted particles at the bottom of the arc. The striker particle is held by a door attached to a track along the arc. The position of the door can be moved along the track in order to achieve different impacting velocities when released. The door is spring-loaded and is released by a solenoid. Once released, particle 1 collides with particle 2, and particles 2 and 3 travel up the arc. Due to gravity, $g$, the particles will eventually come back down the arc and collide a second time, etc.; however, data are only taken before and after the first three-body collision, since the liquid-layer thickness for subsequent collisions cannot be determined as accurately as for this first series. Figure 4 contains a single snapshot taken shortly after the collision for two different cases: (a) a smaller impact velocity that leads to a RNC outcome, and (b) a larger impact velocity that leads to a FS state. The corresponding pre- and post-impact velocities are also plotted as functions of time; the details of these measurements are described below.

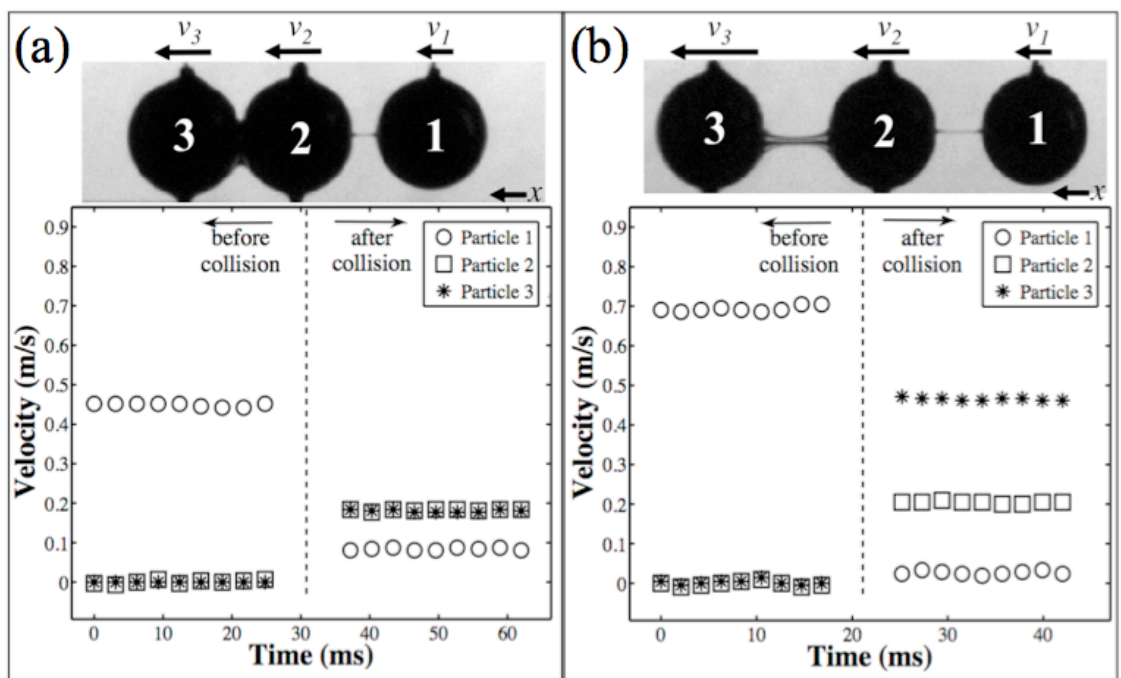

FIGURE 4. Snapshots after collision and corresponding velocity versus time plots for outcomes of (a) RNC and (b) FS using 12000 cP oil viscosity and stainless-steel ball material (case l$\mu$_ss_tn in table 2). The initial velocity of particle 1 is from right to left.

    The particle positions versus time and, hence, velocities of each particle before and after collision are measured using a high-speed camera. The camera is manufactured by DVC (model 340M) with a 640×480 pixel resolution. To increase the rate of image

collection, unnecessary border pixels are cropped out. Depending on the exact distance of the camera and velocity of the striker particle, the resulting resolution is approximately 400×50 to 600×150 pixels. Similar to the high-resolution camera, a Navitar 7000 zoom lens is used so that the camera can be placed approximately 1.5 m away from the pendulum apparatus and wide-angle effects are essentially eliminated. The high-speed camera operates at 40MHz and produces a snapshot every 3.1 ms. The series of snapshots is imported into Matlab to find the position of each particle center in each frame. The grayscale frames are converted into black-and-white images, with white particles appearing on a black background. The particle edges are then eroded using a pre-existing function in Matlab, *imerode*, to separate touching particles so they do not appear to be one object in Matlab. The function *regionprops* calculates the centroid of each particle. Five images before and after the collision are used to calculate the pre- and post-collisional velocities, respectively. The frames immediately before and after the collision, however, are not used due to noise resulting from the collision. The velocities are determined by finding the slope of a linear fit of the centroids of the particles versus time for a given set of five images. The error of the velocity measurement is approximately 0.005 m/s. To verify these measurements, collisions between two dry particles were performed and compared to those performed by Stevens & Hrenya (2005), in which a different measurement technique was used (light-based gates) to measure pre- and post-collisional velocities. The two methods show excellent agreement.

## 3. Theoretical Development

The ultimate objective for a theory describing three-particle, wetted collisions is twofold: to predict the correct outcomes (FA, RNC, NC, and/or FS) over a range of experimental parameters, and to accurately predict the post-collisional velocity of each particle. The first objective, which takes the form of a regime map, serves as a good first gauge of the physics incorporated into the theory, while the second objective involves refinement of the important physics identified in the first step. The focus of this work has been on the first objective, since the findings presented below indicate that the physics necessary to predict the outcomes of 3-body collisions go beyond that previously reported for 2-body collisions. In particular, two physical mechanisms are found to be essential: (i) consideration of the "excess liquid" from the liquid bridge between the initially-agglomerated, target particles (particles 2 and 3); this excess liquid provides additional resistance as the particles separate after collision, and (ii) consideration of the glass-transition (of the oil layer) as a point of rebound due to large lubrication pressures that develop for approaching particles.

To achieve the goal of predicting the correct outcomes, an approximate model is used where a three-body collision is modeled as a series of two-body collisions. First, the striker particle (particle 1 in figure 2) collides with the first target particle (particle 2). Then, the first target collides with the last target particle (particle 3). At this point, particle 1 may "catch up" with particle 2 and then 2 may strike 3 again, and so on; correspondingly, any subsequent collisions are considered. In each two-particle collision, the collision is assumed to have an initial separation of $x_0$ and the collision continues until a final separation of $x_f$ is reached or until the relative velocity becomes zero. If the same particles experience any additional collisions, the same initial and final separations are assumed. The justification for using this two-body approximation for purposes of identifying the important underlying physics is twofold: (i) Donahue *et al.* (2008) found that an analogous approximation predicts well the outcome of three-body collisions between dry particles, though some quantitative improvement in the prediction of post-collisional velocities is obtained using a three-body treatment, and (ii) in-house, preliminary results for a three-body treatment of wet systems indicate that the more complex treatment leads to modest quantitative changes though does not appreciably change the predicted outcomes (i.e., regime map).

### 3.1 *Dimensionless Arguments and Dominant Mechanisms*

The first task in the theoretical development is to identify the predominant mechanisms that govern the behavior. Accordingly, the appropriate dimensionless quantities are assessed. The Reynolds number, $Re$, the capillary number, $Ca$, and the particle Stokes number, $St_{part}$, are calculated for the collisions over the range in the experimental parameter space. Here, $St_{part}$ characterizes the particle inertia as it moves through the surrounding air as opposed to the $St$ defined in equation 1 above, which characterizes forces from particle inertia relative to lubrication forces in the liquid gap. The largest $Re$ encountered experimentally is

$$Re = \rho |v| x / \mu < 0.06,$$

where $\rho$ is the liquid density, $v$ is the relative velocity of the center of particle masses (i.e. $v_1$-$v_2$ or $v_2$-$v_3$), and $x$ is the minimum separation distance between the particles. Since the collisions occur with a low $Re$, Stokes flow prevails in the liquid gap. Additionally, the smallest experimental $Ca$ (ratio of the viscous force to the capillary force) is

$$Ca = 3\mu \bar{a} v / \sigma x > 3400,$$

where $\sigma$ is the surface tension of the silicon oil measured to be 2.4 N/m$^2$. Since the viscous forces dominate, the capillary forces may be neglected. The calculation of $Ca$ is based upon the initial relative velocity of the particles. Finally, $St_{part}$ is always much greater than unity; therefore, the surrounding air medium has negligible effect on the collision dynamics.

### 3.2. *Dynamics of Two-body Wet Collisions*

To describe the Stokes (low $Re$) flow between spheres dominated by viscous forces, a scaling approach is utilized instead of a formal coupling as carried out by Kantak & Davis (2006). Namely, the hydrodynamic equations for undeformed spheres are solved until a rebound criterion is met, which is based upon a scaling argument. This approximation is used, since the goal here is to obtain qualitative agreement with the regime map rather than refining to achieve quantitative agreement, and a formal coupling between the three bodies introduces considerable complexities (i.e., system of coupled, nonlinear, partial differential equations). The kinematic equations describing the hydrodynamic motion of the two particles during a two-body, wet collision are

$$\frac{dx}{dt} = -v(t) \tag{2}$$

and

$$\bar{m}\frac{dv}{dt} = -F_L(t), \tag{3}$$

where $F_L(t)$ is the viscous (lubrication) force resisting the relative motion of the particles in the normal direction. For small deformations and for $x << a$, this force is derived by Kantak & Davis (2006) as

$$F_L(t) = \frac{6\pi\mu\bar{a}^2 v}{x}\left[1 - \frac{x}{x_{\max}}\right]^2,  \tag{4}$$

where $x_{\max}$ is the maximum thickness of the liquid layer. In previous two-body theories, $x_{\max}$ is assumed equal to the initial separation distance for both the approach and rebound stage, but this is not a good assumption for the three-body collisions considered here, as described below. As the particle significantly penetrates far into the liquid layer ($x <<$ $x_{\max}$), the term in the brackets quickly approaches unity and the result for the motion of two immersed spheres moving towards each other is recovered. Using the same assumptions, the absolute pressure in the gap, also derived by Kantak & Davis (2006), is

$$p(r,t) = \frac{3\mu\bar{a}v}{(x + r^2/2\bar{a})^2}\left[1 - \left(\frac{x + r^2/2\bar{a}}{x_{\max}}\right)^2\right] + p_{atm}.  \tag{5}$$

where $r$ is the distance from the axis connecting the two spheres and $p_{atm}$ is the atmospheric pressure. In this work, the pressure is only tracked in $t$ (of which $x$ is a function); therefore, only the maximum pressure between the particles is considered. Here the maximum pressure occurs at the axis of symmetry ($r = 0$). To solve for the relative velocity and separation distance as functions of time, equations 2 and 3 are solved simultaneously using *ode23s* in Matlab, a solver for stiff differential equations. Note that these equations are used to describe particle motion during the approach phase and rebound phase, if encountered (i.e., if an agglomerate is not formed prior to rebound; agglomeration is detected when the relative velocity is equal to zero during the approach or rebound phase). If the rebound criterion is met upon approach, the particles rebound with the relative velocity reversed and multiplied by the dry restitution coefficient, $e_d$, to account for the (kinetic) energy dissipation experienced by the particle during deformation. Specifics on the initial conditions and conditions for reversal of relative velocity (i.e., transition from approach to rebound phase) – the rebound criteria – are detailed below.

### 3.3. *Effect of Excess Fluid in Liquid Bridge*

Upon approach of a given particle pair, the initial separation distance is given by the initial liquid thickness measured using the high-resolution camera described above. The equations above (equations 2 and 3) are solved from this point until conditions meet a rebound criterion that will later be described. If the criterion is met, then the particles begin to rebound until they are separated by a final thickness (unless agglomeration occurs beforehand). In previous two-body work (Davis *et al.* 2002; Ennis *et al.* 1991), the final liquid thickness that the particles encounter upon rebound was assumed equal to the initial (measured) thickness. However, in a three-body collision, the initial target particles (particles 2 and 3 in figure 2) are already in an agglomerated state before the collision. The measured separation distance between particles 2 and 3, $x_{0,2\text{-}3}$ (see figure 3b), characterizes well the "initial" thickness as the particles are approaching each other, but it does not describe well the final liquid thickness experienced by the particles as they rebound until they separate. Since particles 2 and 3 are initially agglomerated (i.e., in contact via their common liquid bridge), "excess" liquid is contained in the bridge (as seen in figure 2 (a)) and serves to fill the widening gap beyond $x_{0,2\text{-}3}$ as the particles separate. More specifically, as the particles separate, the excess liquid will flow in the direction of lowest pressure, which occurs along the centerline ($r = 0$). As a result, the excess liquid in the bridge fills the gap between the separating particles, as illustrated by figure 2 (b). Consequently, the final outbound thickness, $x_{f,2-3}$, is greater than $x_{0,2-3}$ and is related to the amount of excess liquid in the bridge.

Since the measurements of the initial thickness between the target particles (particles 2 and 3) are not adequate to describe the rebound phase of the collision, additional steps must be taken to estimate the "effective" thickness stemming from the excess liquid in the bridge as the particles rebound. The "effective" thickness is intended to be the separation distance at which the particles escape the resistance of the liquid, and not the rupture distance of the liquid bridge as described in Lian, Thornton & Adams (1993), Mikami, Kamiya & Horio (1998), and Pepin, Rossetti & Iveson (2000). Although a small bridge connecting the particles may be present at distances greater than $x_{f,2-3}$, a comparison of the liquid bridge in the high-speed video of the collision and the plots of velocity versus time (such as shown in figure 4) indicates that this bridge provides negligible resistance in the final stages prior to rupture since the velocity remains constant while the bridge is still intact. To calculate $x_{f,2-3}$, the volume of the liquid bridge is divided by the relevant surface area of the particles. In particular, the liquid bridge is approximated as symmetric about the centerline. The shape of the bridge is also approximated to be that of a cylinder ($V_{cyl}$), minus the volume indented by the spherical shape of the particles ($V_{cap}$) at the caps of the cylinder. In this way, $x_{excess,2-3}$ is found by an additional measurement of the height of the liquid bridge ($h$ in figure 2). The volume of the indented cylinder is then calculated as

$$V_{cyl,ind} = V_{cyl} - 2V_{cap}$$
$$= \frac{h^2}{4}\pi\left[ 2\left(a - \sqrt{a^2 - h^2/4}\right) + x_{0,2-3}\right] \tag{6}$$
$$- 2\left\{\frac{1}{3}\pi\left[3a - \left(a - \sqrt{a^2 - h^2/4}\right)\right]\left[a - \sqrt{a^2 - h^2/4}\right]^2\right\}.$$

Assuming that the liquid will be evenly dispersed over the caps as the particles separate, the thickness of the (final) liquid layer between the rebounding particles is the volume of the indented cylinder divided by the area of one cap (dividing by the area of both caps, would only give one-half of the thickness), where the area of one cap is given as

$$A_{cap} = 2\pi a\left(a - \sqrt{a^2 - h^2/4}\right). \tag{7}$$

Accordingly, the liquid-layer thickness upon rebound, when accounting for excess liquid in the bridge, is

$$x_{excess,2-3} = \frac{V_{cyl,ind}}{A_{cap}}$$
$$= \frac{1}{6a}\left[h^2 - 2a\left(a - \sqrt{a^2 - h^2/4}\right) + 3x_{0,2-3}\left(a + \sqrt{a^2 - h^2/4}\right)\right]. \tag{8}$$

The $x_{excess,2-3}$ value calculated in this manner for the experiments is found to be ~1-2 orders of magnitude larger than $x_{0,2-3}$. A similar treatment for the final thickness between particles 1 and 2 is not necessary since the particles are not agglomerated prior to collisions (i.e., no pre-existing liquid bridge is present to provide excess liquid upon rebound). Hence, $x_{f,1-2} = x_{0,1-2}$ for purposes of model calculation.

The calculation of $x_{excess,2-3}$ is a critical component of the model, as can be seen from a comparison of the current model (using $x_{f,2-3} = x_{excess,2-3}$) with predictions from the same model except without considering the bridge using $x_{f,2-3} = x_{0,2-3}$. This treatment of $x_{f,2-3}$ enters the model in two areas: (i) $x_{max}$ in equations 4 and 5 is equal to the largest liquid separation between two particles, so, when considering $x_{f,2-3} = x_{excess,2-3}$, $x_{max}$ is also equal to $x_{excess,2-3}$; and (ii) upon rebound of particles 2 and 3, the differential equations 2 and 3 are solved until the separation of the particles reaches $x_{f,2-3}$. Therefore, if $x_{f,2-3} = x_{0,2-3}$, the equations are solved until a much smaller separation distance is achieved than when $x_{f,2-3} = x_{excess,2-3}$. To illustrate these concepts, figure 5 is a representative plot of the wet restitution coefficient for each particle pair versus $St$. Here, increasing the impact velocity of the striker particle increases $St$, while all other parameters remain unchanged. The wet restitution coefficient between particles 1 and 2 is a ratio of the final velocities over the initial velocities and is defined as

$$e_{w,1-2} = \frac{v_{f,2} - v_{f,1}}{v_{0,1}},$$ (9)

where the subscripts 1 and 2 indicate particles. Similarly, the wet restitution coefficient between particles 2 and 3 is

$$e_{w,2-3} = \frac{v_{f,3} - v_{f,2}}{v_{0,1}},$$ (10)

where it is normalized by the initial velocity of particle 1 since the initial velocities of particles 2 and 3 are zero. When $e_{w,1-2}$ is zero and $e_{w,2-3}$ is zero, the outcome is FA; for $e_{w,1-2}$ zero and $e_{w,2-3}$ non-zero, the outcome is NC; for $e_{w,1-2}$ non-zero and $e_{w,2-3}$ zero, the outcome is RNC; finally, when both are non-zero, the outcome is FS. For collisions between particles that agglomerate, the wet restitution coefficient is zero by definition, and thus the unphysical negative experimental values stem from the error in velocity measurements. In particular, the error in the measurement of the particle velocity propagates to give an error in $e_w$ of approximately 0.02 for low velocities and 0.002 for high velocities. In figure 5a, the thin lines represent the theoretical predictions for $x_{f,2-3} = x_{0,2-3}$, and the thick lines represent the predictions for $x_{f,2-3} = x_{excess,2-3}$. The vertical arrows pointing to $St_{c,1-2}$ and/or $St_{c,2-3}$ are also shown, and the associated outcomes on each side of these values are indicated. The theory without the bridge using $x_{f,2-3} = x_{0,2-3}$ predicts only two outcomes: FA at low $St$ and NC at higher $St$. In contrast, the current model accounting for the excess bridge fluid predicts three outcomes: FA at low $St$, RNC at intermediate $St$, and FS at high $St$. To test the model, figure 5 (b) shows the corresponding experimental data. Then, the data reveal outcomes of FA, RNC, and FS as $St$ increases, in qualitative agreement with the current model and not with the one neglecting the excess bridge fluid. Furthermore, for the model without the bridge using $x_{f,2-3} = x_{0,2-3}$, as the velocity of the striker particle increases, the predicted value of $e_{w,2-3}$ rises rapidly, levels off, and then increases further. The experimental data, on the other hand, indicate that $e_{w,1-2}$ rises rapidly and then decreases before it levels off, and $e_{w,2-3}$ increases smoothly past $St_{c,2-3}$. The same behavior in the experimental data is observed for all of the parameters. In contrast, the current model that utilizes $x_{f,2-3} = x_{excess,2-3}$ qualitatively predicts the correct outcomes (FA, RNC, FS as $St$ increases). Moreover, its features are similar to the experimental results, and the same is true for all of the parameters presented here.

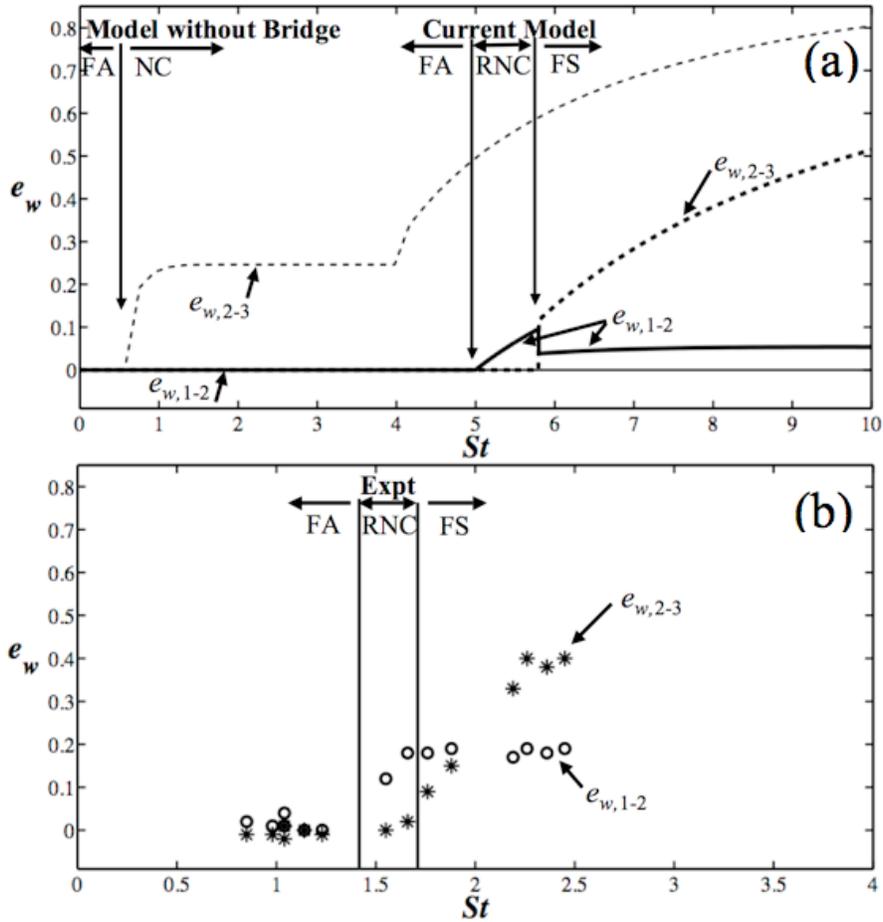

FIGURE 5. Comparisons of (a) theoretical predictions for $e_w$ using the model without the bridge, $x_{f,2\text{-}3} = x_{0,2\text{-}3}$, (thin) and the current model with $x_{f,2\text{-}3} = x_{excess,2\text{-}3}$ (thick), and (b) experimental data using parameters for three-body collisions with 12000 cP viscosity oil, chrome-steel particles and thick oil layer (case hμ_cs_tk in table 2). Both models assume that the oil undergoes a glass transition at $5.5 \times 10^8$ Pa as a rebound criterion, which is the middle of the range of reported for silicon oil. Further details about the glass transition are discussed in section 3.4.

It is important to note that the finding demonstrated in figures 5 (a) and 5 (b), namely that accounting for the effect of the excess liquid in the bridge is crucial in obtaining the correct outcomes, does not stem from an (undue) sensitivity to the input parameters. This concept is illustrated in the regime map of figure 6, which contains a semi-log plot of $x_{f,2\text{-}3}$ versus $St$. In figure 6, the outcomes (FA, NC, RNC, and/or FS) of the collisions have been calculated according to the current model presented above and all parameters are held constant except the final thickness $x_{f,2\text{-}3}$ and impact velocity (which is proportional to $St$). The solid lines indicate the border between regions with different outcomes. The calculated points along the lines are indicated by dots. These lines are slightly jagged due to the discrete nature of the calculated outcomes. This feature could be minimized by greater resolution; however, great computational power would be required. The current computational requirements to create a regime map are significant for two reasons: 1) each three-body collision could contain many two-body collisions (some parts of the parameter space require a large number of collisions, for instance when particles become FA), and 2) the equations are stiff. The dashed lines indicate experimental values of $x_{0,2\text{-}3}$ and $x_{excess,2\text{-}3}$. Consistent with figure 5, more calculations obtained using $x_{f,2\text{-}3} = x_{excess,2\text{-}3}$ predict the ordering of regimes observed experimentally (FA, RNC, FS) whereas predictions obtained using $x_{f,2\text{-}3} = x_{0,2\text{-}3}$ are different (FA, NC). Moreover, it is clear in figure 6 that the erroneous outcomes predicted using $x_{f,2\text{-}3} = x_{0,2\text{-}3}$ do not stem from experimental error, as this value of $x_{f,2\text{-}3}$ is two orders of magnitude smaller than that associated with the correct regimes. Similar to the results depicted in figures 5 (a) and 6, results from the rest of the parameter space also point to the need of accounting for the excess liquid in the bridge between the target particles (i.e., $x_{f,2\text{-}3} = x_{excess,2\text{-}3}$).

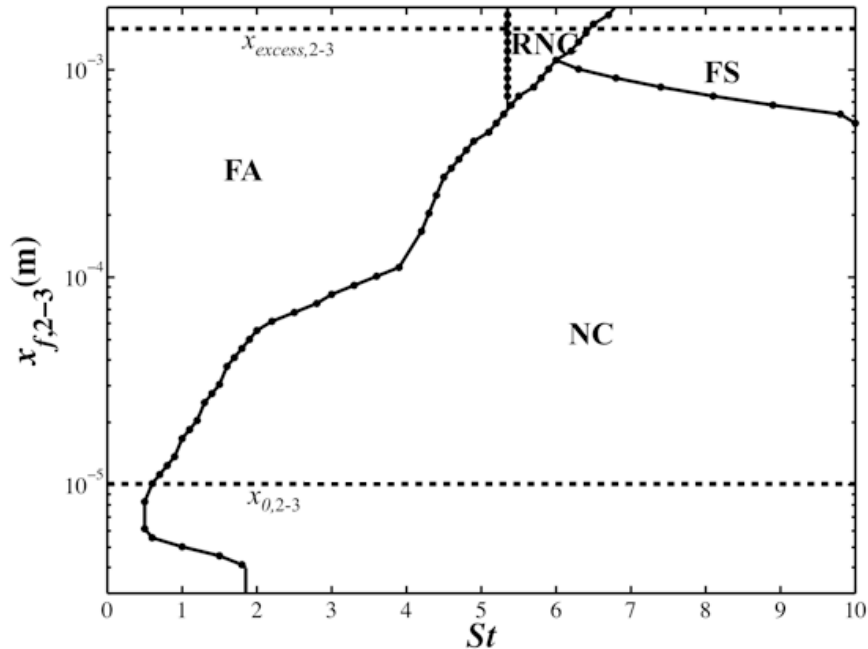

FIGURE 6. Predicted regime map as a function of $x_{f,2\text{-}3}$ and $St$ using parameters for 12000 cP oil and chrome-steel particles and thicker initial thickness (corresponding to case hμ_cs_tk in table 2). Dashed horizontal lines represent $x_{f,2\text{-}3} = x_{excess,2\text{-}3}$ and $x_{f,2\text{-}3} = x_{0,2\text{-}3}$. The model assumes that the oil undergoes a glass transition at $5.5 \times 10^8$ Pa.

### 3.4. Pressure-dependent Glass-Transition

In addition to the excess liquid in the existing bridge between particles 2 and 3, the effect of the pressure on the properties of the oil in the gap is found to be a critical physical process during the three-body collisions. Note that Barnocky & Davis (1989) included pressure dependence in the viscosity of the oil for their work on two-body collisions, though they concluded that its effect was weak for their parameter space. In this work, only the point of glass-transition is considered. The glass transition can be viewed as a simplified way to treat a pressure-dependent viscosity, where the viscosity of the oil is equal to the ambient viscosity at pressures lower than the glass-transition pressure (the pressure at which the silicon oil behaves as a solid), and the viscosity of the oil is infinite at pressures above the glass-transition pressure. For this treatment, the viscosity remains constant throughout the collision process, and rebound will occur if the pressure in the gap reaches the glass-transition pressure. An associated length scale, $x_{gt}$, can be found by letting $r = 0$ and rearranging equation 5 so that $x = x_{gt}$ when $p = p_{gt}$; therefore,

$$x_{gt} = \sqrt{\frac{3\mu\tilde{a}v x_{max}^2}{(p_{gt} - p_{atm})x_{max}^2 + 3\mu\tilde{a}v}} \, . \tag{11}$$

This criterion for rebound is used in addition to those previously used by Davis *et al.* (2002), as described in the next paragraph. In the literature, the glass-transition pressure for silicon oil is reported over a range from $4\times10^8$ Pa (Bair 2008) to $7\times10^8$ Pa (Angel *et al.* 2007).

In previous work by Davis *et al.* (2002), rebound occurs if one of two conditions is met; namely, the particles have sufficient inertia during the collision to penetrate the liquid layer until their separation decreases to an elastohydrodynamic length scale or to the characteristic roughness of the particles. The elastohydrodynamic length scale for rebound is defined as

$$x_r = \left(3\pi\theta\mu\tilde{a}^{3/2}v_0 / \sqrt{2}\right)^{2/5} , \tag{12}$$

where $v_0$ is the initial relative velocity of a given particle-pair collision. Here, $\theta$ is calculated from the material properties of the dry particles and is

$$\theta = \frac{2(1 - \nu^2)}{\pi E^2} \, . \tag{13}$$

The length scale $x_r$ was derived by Davis *et al.* (2002) via a scaling argument, which incorporated the effects of lubrication and elastic theories (i.e., elastohydrodynamics). A more formal treatment of elastohydrodynamics (coupling of equations governing lubrication and particle deformation) was utilized by Kantak & Davis (2006). However, since they assume that cavitation occurs upon rebound, no resistance upon rebound is included in their model. As described above in the context of 3-particle collisions, it is necessary to have resistance upon the rebound, or else the excess liquid from the pre-existing liquid bridge between the target particles would not affect the dynamics. Without rebound resistance, NC would be predicted as one of the outcomes for the experimental parameters (whereas NC never occurred in the experimental parameter space employed) because the rebound resistance between particles 1 and 2 would be much greater than that between 2 and 3 since $x_{0,1\text{-}2} >> x_{0,2\text{-}3}$ in our experiment. Hence, the approximate model of Davis *et al.* (2002) is modified in this effort to include outbound resistance and rebound upon the glass transition. Including the glass transition in the model is an improvement upon Davis *et al.* (2002), since it is unphysical for the particles to continue their approach once the glass-transition pressure has been achieved and even higher pressures would be achieved if particles were allowed to continue their approach.

In the model presented here, the differential equations 2 and 3 are solved from the initial separation until the particle separation decreases to one of three length scales: (i) $x_{gt}$, given in equation 11 (ii), $x_r$, given in equation 12, or (iii) the roughness (bump) size of the particles, $x_b$. In this work, $x_b$ is assumed to be 1 μm based on previous measurements of similar materials (Barnocky & Davis, 1988). For the parameter space examined here (corresponding to the experimental conditions), however, $x_{gt}$ is always encountered before $x_r$ or $x_b$, and so the glass-transition pressure serves as the criterion for rebound.

Given that the glass-transition criterion is not specific to three-body collisions, it is instructive to first compare the various theories for two-particle collisions, since previous theories have shown reasonable agreement with experimental data. In order to clarify the difference among the theories, table 1 is a summary of the wetted two-body models compared here. The heading 'coupling' refers to the coupling of the hydrodynamics and deformation (i.e., how elastohydrodynamics is accounted for); 'scaling' refers to an approximate coupling through the use of $x_r$ as a rebound criterion (as $x_b$ and $x_{gt}$ do not depend on particle material properties), whereas 'formal' refers to the fully-coupled solution of the lubrication equations and deformation equations. For a more direct comparison with the current theory, the Davis *et al.* (2002) theory has been modified in three ways: (i) equations 4 and 5 are solved upon approach and rebound, instead of solving the equations of an immersed sphere where the initial separation in multiplied by 2/3 to account for wetting by the finite larger thickness, (ii) $x_r$ directly depends on the relative velocity as a function of time (therefore $x_r = (6\pi\theta\mu\bar{a}^{3/2}v/\sqrt{2})^{2/5})$, and (iii) outbound resistance is included in the model. In the current model and the modified model of Davis *et al.* (2002), the relative velocity and separation gap are determined using lubrication resistance for undeformed spheres until the gap decreases to the largest of $x_r$, or $x_b$ (Davis *et al.*, 2002) or $x_{gt}$, $x_r$, or $x_b$ (current model), at which point rebound occurs. In Kantak & Davis (2006) the fully-coupled lubrication and elastic deformation equations are solved.

TABLE 1. Two-body wetted model comparisons.

| Model | Coupling | Outbound Resistance | Gap at Which Rebound Occurs |
|---|---|---|---|
| Modified Davis *et al.* (2002) | scaling | yes | largest of $x_r$, $x_b$ |
| Kantak & Davis (2006) | formal | no | variable |
| Current Model | scaling | yes | largest of $x_{gt}$, $x_r$, $x_b$ |

To investigate how well the theories in table 1 perform, figure 7 is two plots of $e_w$ versus $St$ for two-particle collisions with two different viscosities. The wet restitution coefficient for a two-particle collision is defined as

$$e_w = \frac{v_{f,2} - v_{f,1}}{v_{0,1}}, \tag{14}$$

where the subscripts 1 and 2 refer to the striker and target particles, respectively. When $e_w$ is zero, the two particles agglomerate, and when $e_w$ is non-zero, the two particles bounce or separate. Here, experimental data obtained from the Stokes' cradle for 2-particle collisions (points) are compared to the three theories described above. (In the 2-particle implementation of the Stokes' cradle, the striker particle is dry and the single target particle is wetted via the coating bath.) The modified Davis *et al.* (2002) model (thin-dashed-dotted) predicts a larger critical Stokes number, $St_c$, than observed experimentally and under predicts $e_w$ (for non-agglomerated particles) compared to the experimental results. While Kantak & Davis (2006) (thick-dashed) does a good job of predicting $St_c$, it too consistently under predicts $e_w$. Kantak & Davis (2006) also assumes no resistance on rebound; the inclusion of such resistance would shift their predictions to the right on the plot, resulting in a greater mismatch of $St_c$. As mentioned previously, outbound resistance is necessary in order capture the correct outcomes via incorporation of $x_{excess,2\text{-}3}$. Furthermore, the same under predictions of $e_w$ may be seen when compared to their own experimental data (see figure 7 in original article), since the only experimental data presented used particles with $e_d = 0.7$ and yet the theory assumes perfectly elastic particles. The current model (solid) includes an assumed glass-transition pressure of $5.5 \times 10^8$ Pa, in the middle of the range of the pressures reported in the literature. The over prediction of $St_c$ in two-body collisions by the current model is due to the approximate scaling model employed and the treatment of the glass transition, both of which also lead to an over prediction of $St_{c,1\text{-}2}$ and $St_{c,2\text{-}3}$ in three-body collisions. Therefore, a discussion of the over predictions can be found below in section 5 with respect to three-body collisions. The current model makes improvements over its modified predecessor Davis *et al.* (2002) in that it predicts a lower $St_c$ and a higher slope of $e_w$ more consistent with the experimental data. Additionally, the current model also offers some quantitative improvements over Kantak & Davis (2006) in regions of higher $St$ when the current model exhibits a larger $e_w$. Nevertheless, the current model is shifted toward higher $St$ than the observed experimentally. Thus, quantitative difference may be due to the approximate nature of the model and the possibility that there is only partial resistance during the rebound stage of the experiments (such as would be the case if cavitation occurred but only over a portion of the domain or with a dynamic delay).

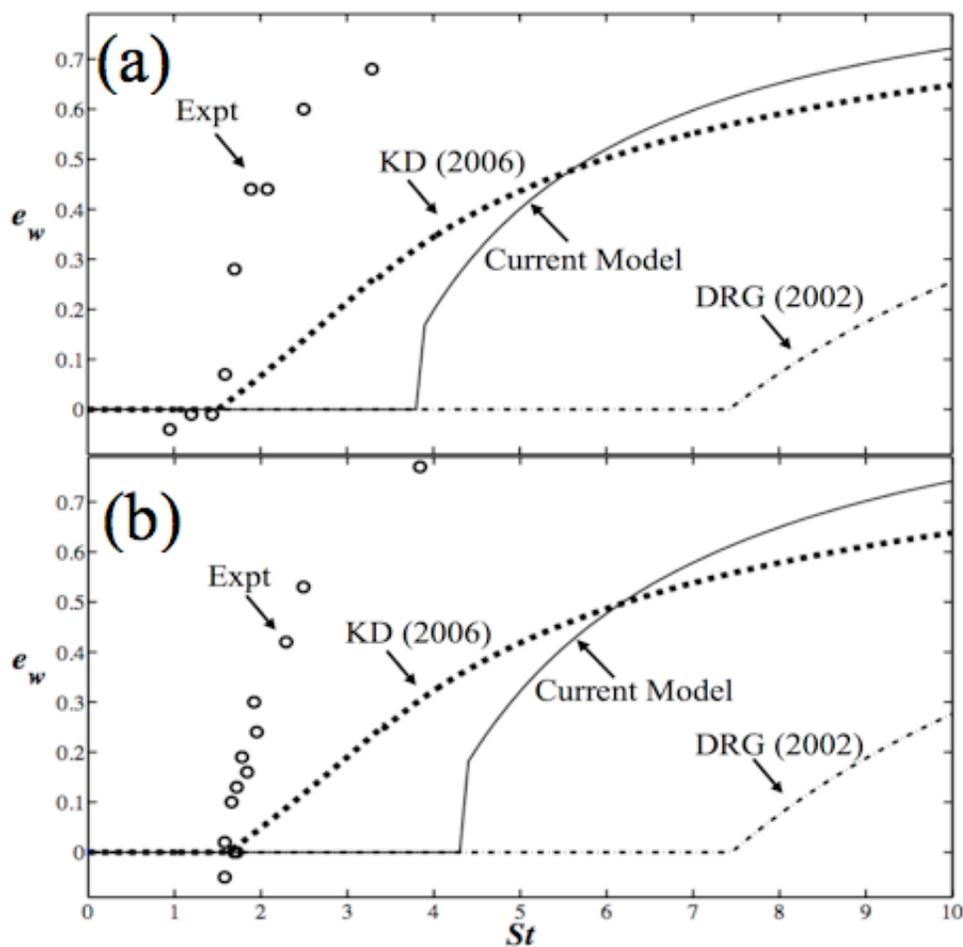

FIGURE 7. Wet restitution coefficient versus Stokes number for wet collisions between two particles with properties of (a) stainless-steel particles, 12000 cP viscosity oil, and 294 mm oil thickness and (b) chrome-steel particles, 5120 cP viscosity oil, and 180 mm oil thickness. Comparisons are presented between experimental results and theories proposed by a modified form of Davis *et al.* (2002) [DRG (2002)], Kantak and Davis (2006) [KD (2006)] and the current model using a glass-transition pressure of $5.5 \times 10^8$ Pa.

The improvement that the inclusion of the glass-transition criterion for rebound makes relative to Davis *et al.* (2002) for two-particle collisions is found to be crucial in predicting the correct outcomes of three-body collisions. In figure 8, the three-body collisions are modeled as a series of two-body collisions. The thin lines represent the modified theory of Davis *et al.* (2002) without considering the glass-transition. The thick lines represent the current theory that includes the condition of rebound at the glass-transition pressure of $5.5 \times 10^8$ Pa. The vertical arrows demarcate the outcomes for an easy comparison. As seen in both figure 8 (a) and figure 8 (b) for the two viscosities, the experimental outcomes observed as *St* increases (for the range of *St* examined) are FA, RNC, and FS, respectively. The predictions using the model of Davis *et al.* (2002) produces outcomes of FA and NC for 12000 cP, and FA, RNC, NC for 5120 cP. For the current theory, which has the glass-transition pressure as a rebound condition, the outcomes for 12000 cP are in qualitative agreement with experiment. However, the outcomes predicted for 5120 cP viscosity are FA, RNC, NC and FS, which differ from experimental outcomes since NC was not observed. For the plots using 5120 cP viscosity, within the region of RNC, $e_{w,1\text{-}2}$ is relatively small, as it is for all 5120 cP plots presented in this work. Similar to the two-particle collisions (figure 8), the approximate theories over-predict the observed critical Stokes numbers. As mentioned previously, the $Ca$ is based upon the *initial* relative velocity of the particles. Because the collisions of particles with 5120 cP oil have small *final* relative velocities between particles 1 and 2, it is worthwhile to revisit the assumption of neglected capillary forces to determine whether or not the RNC region, which occurs over a very small range of $St$, is still predicted. More specifically, if capillary forces are considered in this region, RNC may not be predicted since particles 1 and 2 would be more likely to agglomerate due to the additional cohesion associated with capillary forces. However, the $Ca$ is found to still be substantially greater than unity between particles 1 and 2 for $St$ within the region where RNC is predicted, when using the final relative velocities of the particles (rather than initial). Therefore, even if capillary forces were considered here, RNC would be still predicted and the predicted progression of outcomes for all parameters explored would remain the same.

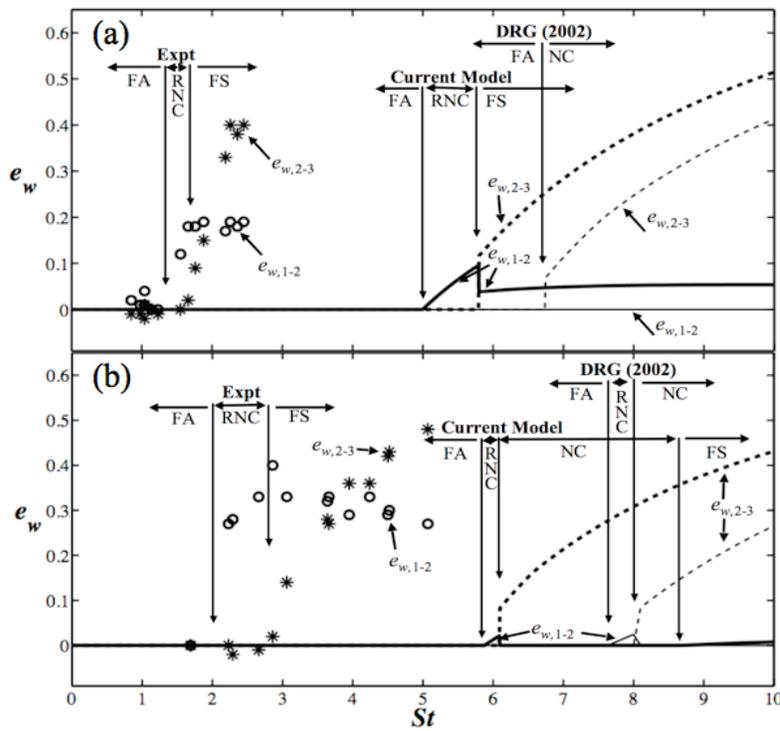

FIGURE 8. Wet restitution coefficient versus Stokes number for normal three-particle collisions with (a) 12000 cP oil viscosity and stainless-steel particles (case hμ_cs_tk in table 2), and (b) 5120 cP oil viscosity and chrome-steel particles (case lμ_ss_tn in table 2). The experimental results are compared against the modified theory of Davis *et al.* (2002) [DRG (2002)], represented by the thin lines, and the current model that uses the glass-transition pressure equal to $5.5 \times 10^{8}$ Pa as a rebound point, represented by the thick lines.

   To gain insight into the source of the additional predicted outcome (NC) relative to that observed experimentally for 5120 cP viscosity (figure 8 (b)), a regime map of the predicted outcomes as a function of the glass-transition pressure and $St$ is plotted in figure 9. The dashed lines represent the reported glass-transition pressures. In this work, $5.5 \times 10^8$ Pa has been used for model predictions, since it is the midpoint of the reported values. However, the regime map (figure 9) clearly indicates that the predicted outcomes, over this range of glass-transition pressures, are near a transitional point on the regime map. For instance, a glass-transition pressure of $3 \times 10^8$ Pa predicts the correct outcomes, which is fairly close to the reported range, especially considering the width of the reported range. Consequently, the experimental/model mismatch does not provide enough evidence of the need for an improvement of the overall physics, only a refinement of the approximations.

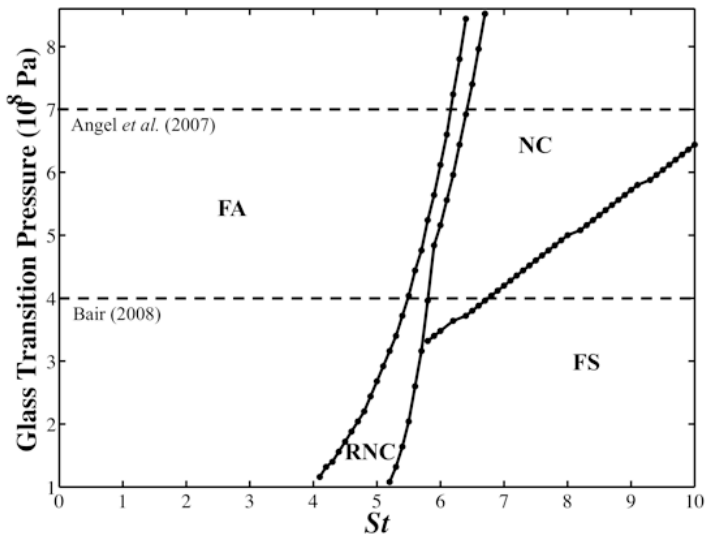

FIGURE 9. Regime map of glass-transition pressure versus $St$ for 5120 cP viscosity oil, chrome-steel balls, and thinner (case lµ_cs_tn in table 2). The dashed lines demarcate the range of the glass-transition pressure for silicon oil that has been reported.

### 3.5. *Model Summary*

   To recap, the theory for 3-body collisions that has been developed in this section expands upon the two-body, scaling theory derived by Davis *et al.* (2002). In particular, the position and velocities of the particles are found by considering the three-body collision as a series of two-body collisions and solving the kinematic equations above (equations 2 and 3) for each collision. In contrast to previous works, here the value of the maximum liquid-layer thickness, $x_{max}$, for the collision between the initial agglomerated targets in equations 4 and 5 is changed to equal $x_{excess,2-3}$ due to a pre-existing liquid bridge (not present in two-particle systems). Equations 2 and 3 are solved with an initial separation equal to the initial (measured) thickness. They are solved for decreasing separation of the sphere noses during the approach stage until one of three rebound criteria is met, two of which were previously explored in Davis *et al.* (2002), namely, the separation distance decreases to $x_b$, $x_r$ or $x_{gt}$, where surface roughness, elastic deformation, or the glass transition, respectively, becomes important. The additional (third) rebound condition used here is the length scale that incorporates the effects of the glass transition, which for the parameter space explored here, is always encountered before the other two conditions. At the beginning of the rebound stage, the relative velocity is equal to the negative velocity at the time at which the rebound condition was achieved multiplied by $e_d$. The kinematic equations are again solved until the gap between the particles increases to $x_f$, at which point separation occurs. For the collision between particles 1 and 2, $x_{f,1\text{-}2}$ remains equal to $x_{0,1\text{-}2}$; between particles 2 and 3, $x_{f,2\text{-}3}$ is now equal to $x_{excess,2\text{-}3}$, since these particles are agglomerated before the collision and their liquid bridge contributes excess liquid to the gap as the spheres separate. If at any time during this process the relative velocity equals zero, agglomeration occurs and any further integration in time is not required.

## 4. Additional Results and Discussion

Now that the important physics of three-particle collisions have been identified, the objective of the current section is twofold: (i) to further assess the ability of the model to predict the correct progression of outcomes over a wider range of experimental parameters, and (ii) to determine the ability of the model to predict trends in the plot of $e_w$ versus $St$ as experimental parameters are varied. For all cases, model predictions are obtained using the theory described above, namely via an approximation of the three-body collision as a series of two-body collisions, using an effective thickness based upon the excess liquid in the bridge as a final thickness between the target particles, and adding glass-transition effects as a condition of rebound.

With regard to the first objective, a listing of the varied experimental parameters is found in table 2 along with the corresponding outcomes, both experimental and predicted, in order of increasing impact velocity (or, equivalently, increasing $St$). Parameters that are varied include: oil viscosity, particle material, oil thicknesses (including $x_{0,1\text{-}2}$, $x_{0,2\text{-}3}$ and $x_{excess,2\text{-}3}$), and impact velocity. The notation used to describe each case refers to viscosity, high (hμ) or low (lμ); particle material, chrome steel (cs) or stainless steel (ss); and liquid thickness, thick (tk) or thin (tn). Various oil thicknesses are achieved by varying the drip time (i.e., time to collision after immersion in the coating bath) as illustrated in figure 3 (c). The particles drip for either 60 (thick) or 120 (thin) seconds before a collision. The experimental outcomes in all cases are FA, RNC and FS as the impact velocity is increased. For all three-particle collisions involving the higher-viscosity 12000 cP silicon oil (cases hμ_cs_tk − hμ_ss_tn), the outcomes predicted are the same as the outcomes observed experimentally. In the collisions involving the lower-viscosity 5120 cP silicon oil (cases lμ_cs_tk − lμ_ss_tn), the predicted outcomes contain all of the observed outcomes in the correct order, the only difference being that an additional outcome of NC is predicted. However, as described in the section above and illustrated in figure 7, this discrepancy can be explained via the proximity of the predictions to a transitional point on the regime map and uncertainty in previous measurements of the glass-transition pressure, as well as the approximate nature of the scaling theory.

TABLE 2. Experimental parameters for normal, wetted, 3-particle collisions. Experimental and predicted outcomes are listed in order of increasing velocity of the striker particle. The possible outcomes are fully agglomerated (FA), Newton's cradle (NC), reverse Newton's cradle (RNC), and fully separated (FS). A glass-transition pressure of $5.5 \times 10^8$ Pa is used for all predictions.

| case | viscosity (cP) | steel material | particle material | $x_{0,1\text{-}2}$ (µm) | $x_{0,2\text{-}3}$ (µm) | $x_{\text{excess},2\text{-}3}$ (µm) | impact velocities (m/s) | experimental outcomes | predicted outcomes |
|---|---|---|---|---|---|---|---|---|---|
| hµ_cs_tk | 12000 | chrome | | 412 | 10 | 1534 | $0.22 - 0.65$ | FA, RNC, FS | FA, RNC, FS |
| hµ_cs_tn | 12000 | chrome | | 310 | 7 | 1138 | $0.23 - 0.83$ | FA, RNC, FS | FA, RNC, FS |
| hµ_ss_tk | 12000 | stainless | | 416 | 10 | 1577 | $0.11 - 1.8$ | FA, RNC, FS | FA, RNC, FS |
| hµ_ss_tn | 12000 | stainless | | 313 | 7 | 1155 | $0.11 - 0.84$ | FA, RNC, FS | FA, RNC, FS |
| lµ_cs_tk | 5120 | chrome | | 280 | 10 | 1106 | $0.14 - 0.55$ | FA, RNC, FS | FA, NC, RNC, FS |
| lµ_cs_tn | 5120 | chrome | | 202 | 10 | 892 | $0.17 - 0.87$ | FA, RNC, FS | FA, NC, RNC, FS |
| lµ_ss_tk | 5120 | stainless | | 295 | 10 | 1105 | $0.11 - 0.40$ | FA, RNC, FS | FA, NC, RNC, FS |
| lµ_ss_tn | 5120 | stainless | | 219 | 10 | 890 | $0.19 - 0.51$ | FA, RNC, FS | FA, NC, RNC, FS |

Related to the second objective mentioned above, the theory is able to predict the same trends in $St_{c,1\text{-}2}$ and $St_{c,2\text{-}3}$ as the experimental parameters are varied. First, the viscosity of the oil is investigated. To show robustness, figure 10 is a plot of $e_w$ versus $St$ for both (a) chrome steel and (b) stainless steel. The experimental results are shown here as points, but only demarcations of $St_{c,1\text{-}2}$ for the current model are shown for a qualitative comparison. As the viscosity is increased, the experimental results for both $St_{c,1\text{-}2}$ and $St_{c,2\text{-}3}$ decrease (i.e., the particles have a larger tendency to rebound for a given $St$). As shown, the model is in qualitative agreement with these trends. Observing smaller $St_{c,1\text{-}2}$ and $St_{c,2\text{-}3}$ with larger viscosity may at first seem counterintuitive, since a high viscosity implies a "stickier" collision. In particular, if $e_w$ is plotted against the dimensional impact velocity instead of the dimensionless $St$, the lower-viscosity oil would experience a transition from FA to RNC at a smaller impact velocity; therefore, in practice, as viscosity is increased the collision is indeed "stickier" and separation occurs at a higher impact velocities. The predicted trend can be traced to the rebound criteria contained in the model. In previous modeling of two-body collisions by Ennis *et al.* (1991), the only rebound criterion used was surface roughness ($x_b$), and $e_w$ was related to the parameters by

$$e_w = \begin{cases} 0 & , \quad St < St_c \\ e_d(1 - St_c/St), & St > St_c \end{cases} \tag{15}$$

and

$$St_c = \frac{1}{e_d}\ln\left(\frac{x_b}{x_0}\right). \tag{16}$$

Notice that $St_c$ has no dependence on the viscosity, which is contrary to the data contained in figure 10. In contrast, for the work by Kantak & Davis (2006), elastohydrodynamics correctly predicts the decrease in $St_c$ for two particles with increasing viscosity. In their work, the trend stems from the fact that as the pressure increases the particles deform more, leading to a greater storage of energy to be released. Therefore, since the pressure increases more with a larger oil viscosity, there is more deformation of the particles, and the collision has a smaller $St_c$ with larger viscosity. Similarly, this physical process is accounted for in the scaling analysis by Davis *et al.* (2002) since, in equation 12 and 13, $x_r$ depends on the solid-particle properties, namely $E$ and $v$. In the current model, the correct trends are predicted even though the $x_r$ does not serve as the rebound length scale. Instead, the glass-transition length scale, $x_{gt}$, prevails. Accordingly, the point of rebound is only dependent upon the pressure between the particles, not the solid-particle properties. Therefore, in this work the mechanism for the observed trend with viscosity does not arise from elastohydrodynamic theory, but rather from the relation between pressure and viscosity. As seen in equation 5, the pressure is proportional to viscosity, and so a higher pressure is achieved with a higher viscosity. Therefore rebound at the glass-transition pressure can be achieved at a larger separation distance with a high viscosity. This result can be also seen in equation 11, where $x_{gt}$ increases as the viscosity increases.

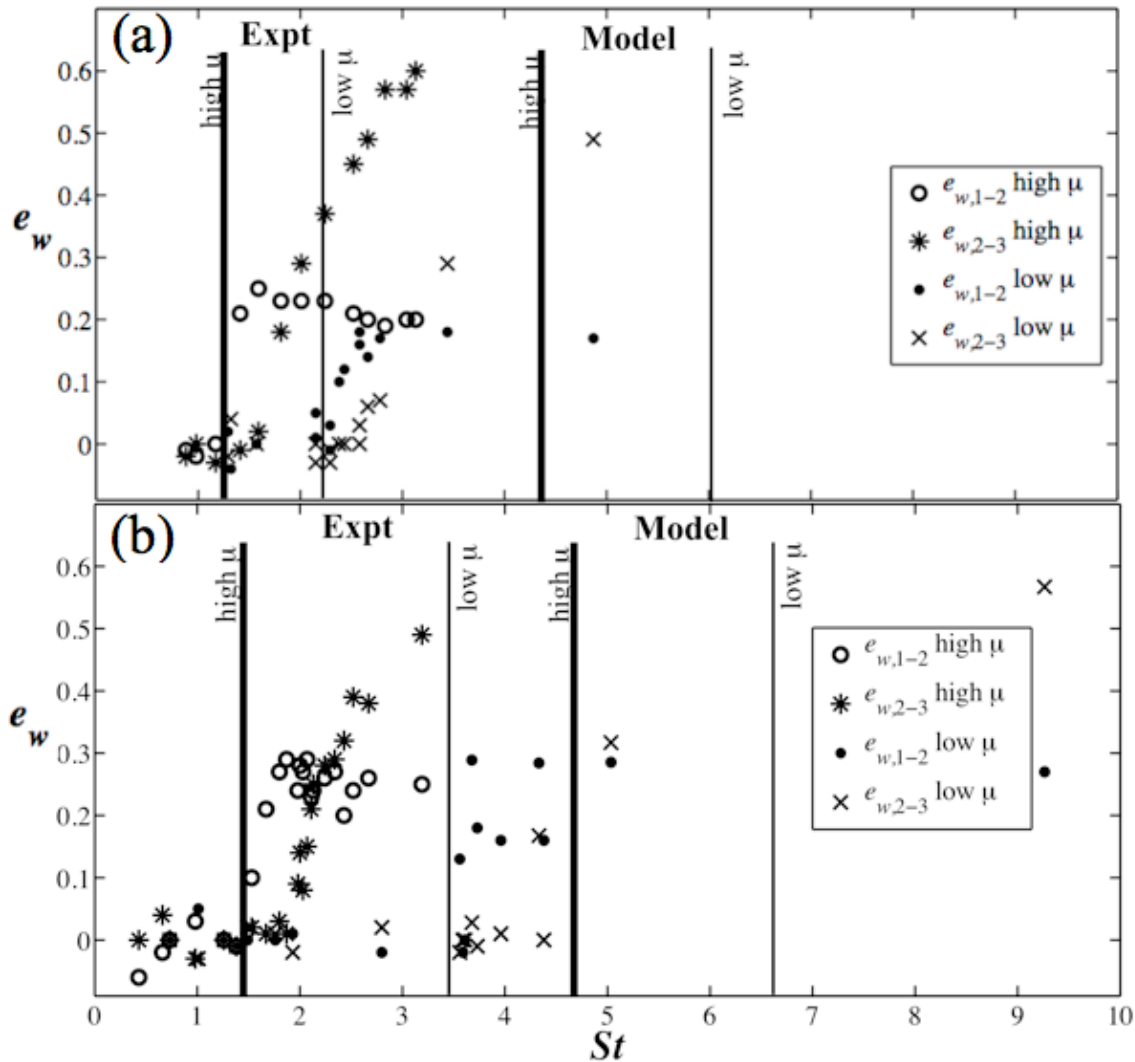

FIGURE 10. Effect of oil viscosity on the wet restitution coefficient for (a) chrome steel and thicker liquid layer (cases hμ_cs_tk and lμ_cs_tk), and (b) stainless steel and thinner liquid layer (cases hμ_ss_tn and lμ_ss_tn). The vertical solid lines demarcate $St_{c,1-2}$ and show that this critical value for rebound shifts to higher values for both theory and experiment as the viscosity is decreased.

Although the glass-transition is not dependent upon the solid-particle properties, these properties do have an impact on the dynamics of the collision upon velocity reversal (via particle deformation). In particular, the influence of the dry restitution coefficient is demonstrated in figure 11, where viscosity and all thicknesses are held constant while varying the two different types of particle material, chrome steel ($e_d$ = 0.99) and stainless steel ($e_d$ = 0.90). Both the experiment and theory agree that, as the dry restitution coefficient increases, $St_{c,1-2}$ and $St_{c,2-3}$ decrease and $e_{w,1-2}$ and $e_{w,2-3}$ increase. As expected, the softer particles will experience a greater energy loss during collisions, and thus are more likely to agglomerate. In the theory, upon rebound, the particles have a relative velocity equal to the negative of the relative velocity when the rebound criterion is met, multiplied by $e_d$. Therefore, after a collision between two particles, a smaller $e_d$ results in a smaller relative velocity and thus a smaller $e_w$. Nonetheless, since the difference in the dry restitution coefficients between the two particle materials is small, the shift seen is also small. Here, $e_d$ refers to the dry restitution coefficient between two steel particles, since measurements are not available between steel and solid silicon oil. As seen from figure 11, this approximation is able to capture the appropriate trends between harder and softer particles.

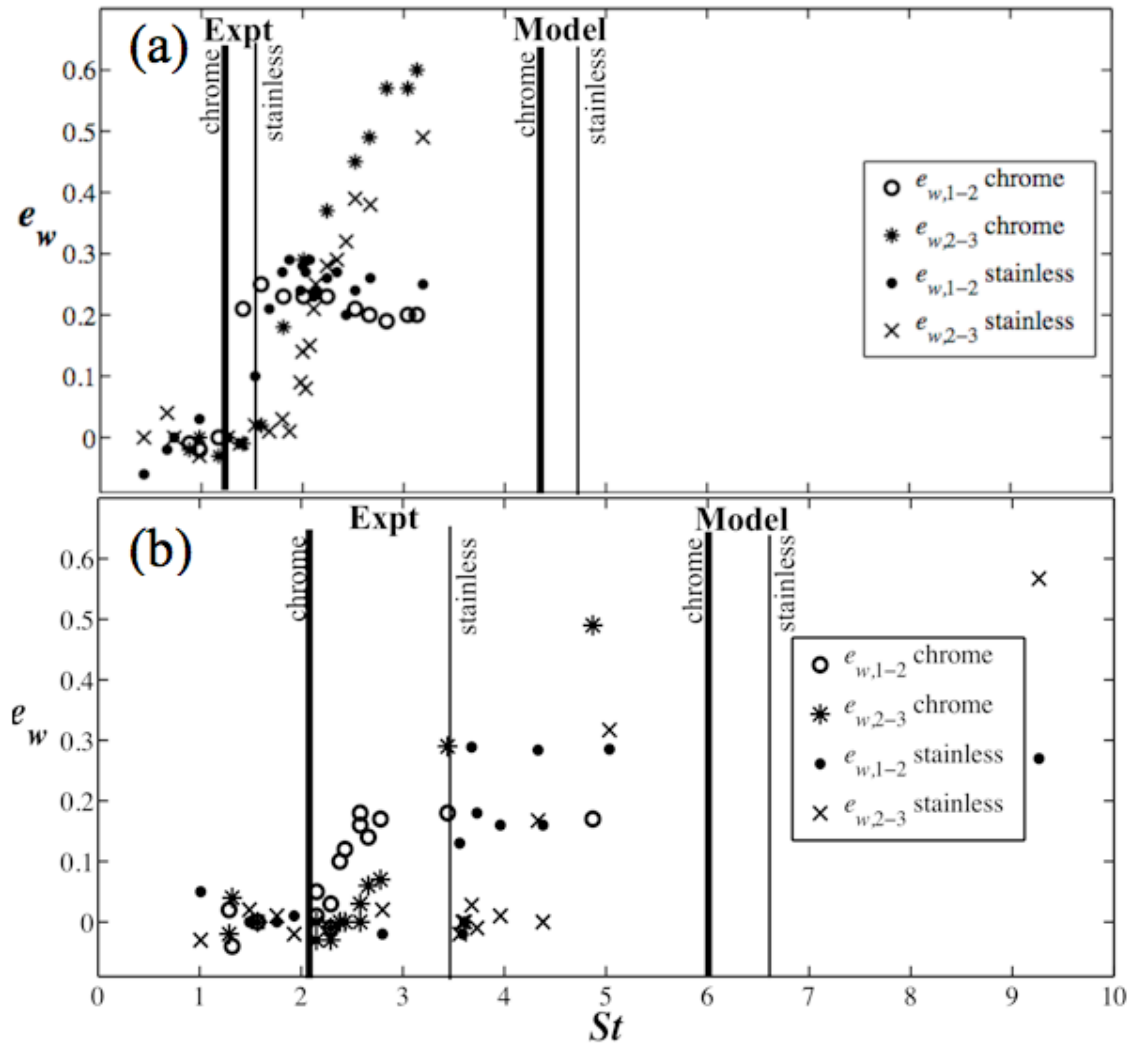

FIGURE 11. Effect of particle material on the wet restitution coefficient with (a) 12000 cP and thicker liquid layer (cases hμ_cs_tk and hμ_ss_tk), and (b) 5120 cP and thinner liquid layer (cases lμ_cs_tn and lμ_ss_tn). The vertical solid lines demarcate $St_{c,1-2}$ and show this critical value for rebound shifts to higher values for both theory and experiment as the dry restitution coefficient is decreased.

Finally, in figure 12, the effect of the liquid-layer thickness on $e_w$ is illustrated. Different liquid-layer thicknesses are achieved by allowing the target particles to drip for a longer period of time. Consequently, all three liquid thicknesses are smaller when the particles are allowed to drip for a longer time. In both figure 12 (a) and figure 12 (b), a qualitative agreement exists between experiment and theory, and a thinner oil layer has a lower $St_{c,1\text{-}2}$ and $St_{c,2\text{-}3}$, and a higher $e_{w,1\text{-}2}$ and $e_{w,2\text{-}3}$. With a thinner oil layer, the particles have a smaller distance to travel during approach to meet a rebound criterion (since none of the rebound criteria depend on oil thickness), and they have a smaller final distance to travel during rebound to become separated. In other words, the resistance to particle motion is decreased, and agglomeration is less likely.

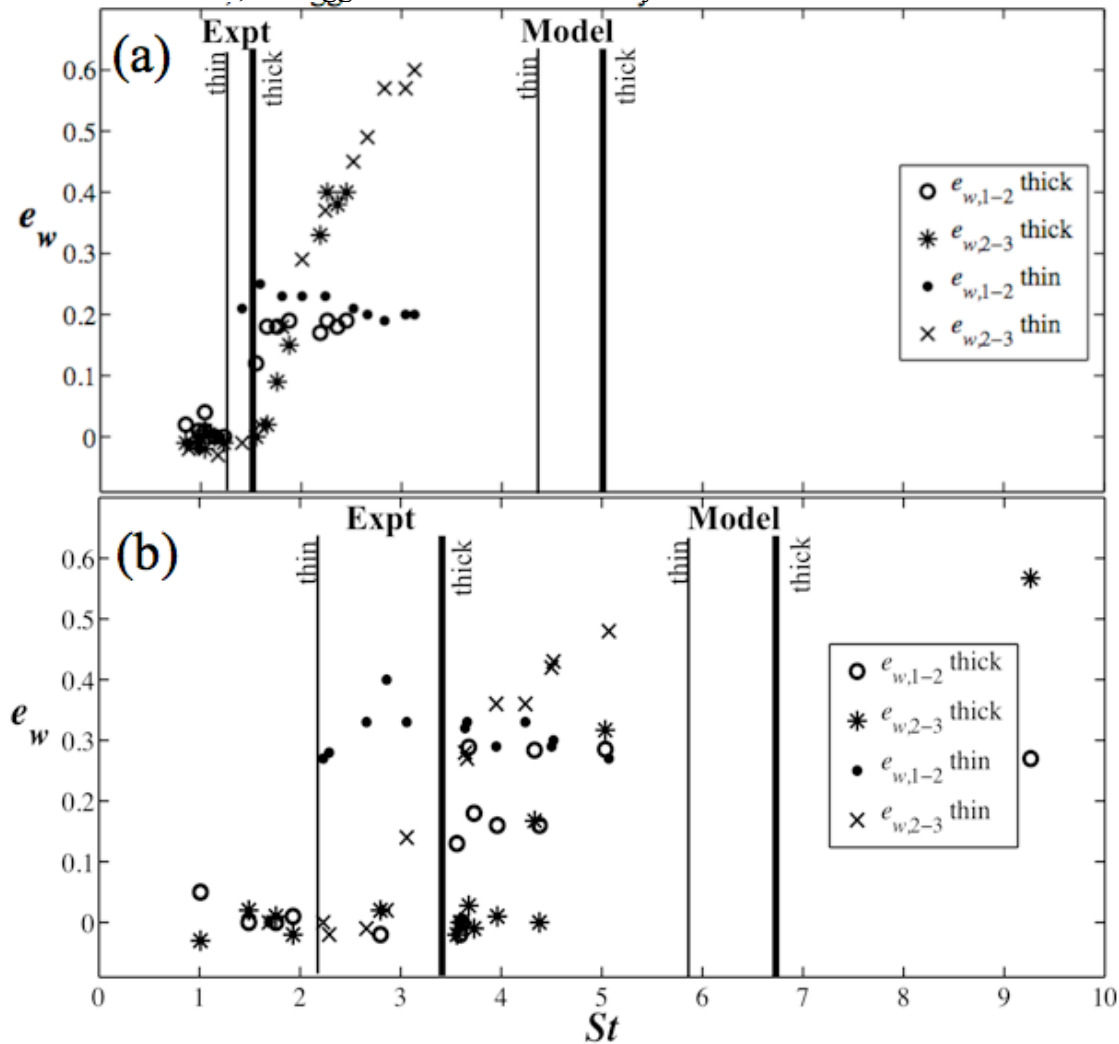

FIGURE 12. Effect of oil thickness on the wet restitution coefficient for (a) 12000 cP oil viscosity, chrome steel (cases hµ_cs_tk and hµ_cs_tn), and (b) 5120 cP, stainless steel (cases lµ_ss_tk and lµ_ss_tn). The vertical solid lines demarcate $St_{c,1\text{-}2}$ and show that this critical value for rebound shifts to lower values for both theory and experiment as liquid-layer thickness is decreased.

## 5. Summary

Unlike previous efforts on collisions between wetted particles (particles with a thin coating of viscous liquid), which focused on two-body systems, the focus of this work is on the dynamics of three-body, wetted collisions. Here, normal or head-on collisions are considered, in which four outcomes are geometrically possible, unlike two-particle collisions in which only two outcomes are possible. To better understand the underlying physics of this three-body system, a combination of experiments and lubrication (low Reynolds number) theory is used.

The experiments are carried out with a Stokes' cradle, which is an apparatus inspired by the desktop toy known as the Newton's cradle. Unlike the Newton's cradle, however, the particles in the Stokes' cradle are wetted prior to collision. Measurements of the liquid-coating thickness and pre- and post-collisional velocities were made using a high-resolution camera and a high-speed camera, respectively. Parameters varied include the oil viscosity, particle material, thicknesses of the oil layer, and the impact velocity. In this work, only outcomes of FA (fully agglomerated), RNC (reverse Newton's cradle) and FS (fully separated) were observed. Surprisingly, the outcome most commonly associated with the desktop toy, NC, proved to be elusive for the conditions investigated. More detail on how investigation of the regime maps lead to experimental realization of NC can be found in Donahue, Hrenya & Davis (2009).

Comparisons of the experimental results are made against theory that approximates the three-particle collision as a series of two-particle collisions. The objective of the modeling is to achieve qualitative agreement with experimental data in order to identify the dominant physical mechanisms at play during the collision. One evaluation of the qualitative results is made by comparing the experimental outcomes with the predicted outcomes. Previous models for wetted, two-body collisions which assume Stokes' flow (low $Re$) and particle deformation, do not result in the correct outcomes for three-body systems, and a regime map of the parameters reveals that the mismatch does not result from a (realistic) sensitivity to the input parameters. Accordingly, a scaling model has been developed here that has two key differences from previous two-body models. First, in a three-particle collision, since the initially agglomerated target particles have a liquid bridge that contains a large amount of "excess" liquid (not found in a two-particle collision), an effective thickness based upon the excess liquid that fills in the gap between the particles as they separate must be incorporated. Second, unlike most previous two-body theories (Davis et al. 1986; Ennis et al. 1991; Davis et al. 2002; Kantak & Davis 2006), a rebound criterion has been developed which ensures rebound as the pressure between the particles reaches the glass-transition pressure (pressure at which the oil behaves as a solid). A good model/experimental qualitative agreement for the outcomes (i.e. FA, RNC, FS) is found when the above physics are taken into consideration.

In addition to predicting the outcomes, the proposed theory also predicts the qualitative trends in $St_{c,1-2}$ and $St_{c,2-3}$ as experimental parameters are varied. Most notably, as the viscosity of the oil is increased, $St_{c,1-2}$ and $St_{c,2-3}$ decrease. Unlike in previous two-body theories, where the same trend arose from elastohydrodynamics, here the glass transition is the source of this behavior. Namely, since the pressure between the particles increases with viscosity (equation 5), higher pressure is obtained with higher viscosity oil. Therefore, the glass-transition pressure is reached at larger separation distances with higher viscosity oil.

Due to the predicted outcomes and trends showing qualitative agreement with the experiments, the important physical processes have been identified. The scaling analysis used is ideal for this process because it helps to quickly identify any gross mismatches without a comprehensive computational effort. An improved model is required for a more accurate quantitative match, and this can be achieved by refinement upon two approximations: (i) simultaneous treatment of the three-body collision rather than the series of two-particle collisions, which is expected to be particularly important for wet collisions since lubrication forces act simultaneously on both sides of the middle ball; and (ii) a strict comprehensive coupling of the hydrodynamic (which includes a pressure-dependent viscosity, stiff in nature) and the elastic theories. For a complete model of collisions occurring in practice, oblique collisions will also need to be considered.

The authors are grateful to Deniz Ertas and Scott Bair for helpful conversations during the course of the project. This work was supported by the National Science Foundation (CBET 0754825) and by National Aeronautical and Space Administration (NNC05GA48G). C.M.D. would also like to acknowledge fellowship assistance from the Department of Education GAANN Program (P200A060265).

## References

Angel R. J., Bujak M, Zhao J, Gatta GD, Jacobsen SD. 2007. Effective hydrostatic limits of pressure media for high-pressure crystallographic studies. *Journal of Applied Crystallography* 40:26-32

Bair S. 2008. *Unpublished*

Barnocky G, Davis RH. 1988. Elastohydrodynamic Collision and Rebound of Spheres - Experimental-Verification. *Physics of Fluids* 31:1324-9

Barnocky G, Davis RH. 1989. The Influence of Pressure-Dependent Density and Viscosity on the Elastohydrodynamic Collision and Rebound of 2 Spheres. *Journal of Fluid Mechanics* 209:501-19

Chu PSY, Cameron J. 1962. Pressure Viscosity Characteristics of Lubricating Oils. *J. Inst. Petrol.* 48:147

Davis RH, Rager DA, Good BT. 2002. Elastohydrodynamic rebound of spheres from coated surfaces. *Journal of Fluid Mechanics* 468:107-19

Davis RH, Serayssol JM, Hinch EJ. 1986. The Elastohydrodynamic Collision of 2 Spheres. *Journal of Fluid Mechanics* 163:479-97

Donahue CM, Hrenya CM, Davis RH. 2009. Stokes' Cradle: Adding a Layer of Complexity to Newton's Cradle. *In Preparation*

Donahue CM, Hrenya CM, Zelinskaya AP, Nakagawa KJ. 2008. Newton's cradle undone: Experiments and collision models for the normal collision of three solid spheres. *Physics of Fluids* 20

Ennis BJ, Tardos G, Pfeffer R. 1991. A Microlevel-Based Characterization of Granulation Phenomena. *Powder Technology* 65:257-72

Gondret P, Lance M, Petit L. 2002. Bouncing motion of spherical particles in fluids. *Physics of Fluids* 14:643-52

Hertz H. 1882. On the Contact of Rigid Elastic Solids and on Hardness. *Journal fur die Reine und Angewandte Mathematik* 94:156-71

Joseph GG, Zenit R, Hunt ML, Rosenwinkel AM. 2001. Particle-wall collisions in a viscous fluid. *Journal of Fluid Mechanics* 433:329-46

Kantak AA, Davis RH. 2006. Elastohydrodynamic theory for wet oblique collisions. *Powder Technology* 168:42-52

Lian G, Adams MJ, Thornton C. 1996. Elastohydrodynamic collisions of solid spheres. *Journal of Fluid Mechanics* 311:141-52

Lian GP, Thornton C, Adams MJ. 1993. A Theoretical-Study of the Liquid Bridge Forces between 2 Rigid Spherical Bodies. *Journal of Colloid and Interface Science* 161:138-47

Lundberg J, Shen HH. 1992. Collisional Restitution Dependence on Viscosity. *Journal of Engineering Mechanics-Asce* 118:979-89

Mikami T, Kamiya H, Horio M. 1998. Numerical simulation of cohesive powder behavior in a fluidized bed. *Chemical Engineering Science* 53:1927-40

Pepin X, Rossetti D, Iveson S. 2000. Modeling the Evolution and Rupture of Pendular Liquid Bridges in the Presence of Large Wetting Hysteresis *Journal of Colloid and Interface Science* 232:289-97

Serayssol JM, Davis RH. 1986. The Influence of Surface Interactions on the Elastohydrodynamic Collision of 2 Spheres. *Journal of Colloid and Interface Science* 114:54-66

Stevens AB, Hrenya CM. 2005. Comparison of soft-sphere models to measurements of collision properties during normal impacts. *Powder Technology* 154:99-109